# Digital Detox – Mitigating Digital Overuse in Times of Remote Work and Social Isolation

## Research-in-progress

**Milad Mirbabaie**
Faculty of Business Studies and Economics
University of Bremen
Bremen, Germany
Email: milad.mirbabaie@uni-bremen.de

**Julian Marx**
Department of Computer Science and Applied Cognitive Science
University of Duisburg-Essen
Duisburg, Germany
Email: julian.marx@uni-due.de

**Lea-Marie Braun**
Department of Computer Science and Applied Cognitive Science
University of Duisburg-Essen
Duisburg, Germany
Email: lea-marie.braun@uni-due.de

**Stefan Stieglitz**
Department of Computer Science and Applied Cognitive Science
University of Duisburg-Essen
Duisburg, Germany
Email: stefan.stieglitz@uni-due.de

## Abstract

Remote work arrangements and limited recreational options in times of social isolation increase the risk of digital overuse for individuals. Its consequences can range from impaired mental health to issues of technology addiction. A conformant countermovement has popularised "digital detoxing", a practice that endorses to deliberately limit technology use to reduce digital involvement and physiological stress. In times of social isolation, however, digital networking may provide the principle access to social interactions. To provide empirical evidence about the sweet spot between mitigating digital overuse and perceived social connectedness, this paper proposes a mixed-methods design to scrutinise the impact of digital detox measures in a professional context. Possible results will help to better understand how digital overuse may effectively be mitigated by remote workers and what measures organisations can take to create a digital environment that supports employee satisfaction and mental health.

# 1   Introduction

Digitisation is changing society in various fields, including the world of work and individuals' private life (Degryse 2016). Therefore, professionals are confronted with technology not only in the context of work, but also during the time enclosing their working hours (Kapidzic et al. 2019). In particular, communication technology has changed the way people plan and conduct their social activities (Techatassanasoontorn and Thaiprasert 2013). It mediates the way individuals communicate with one another and how relationships are built and maintained (Chayko 2014). Hence, communication technology is often regarded as a tool used by individuals to satisfy their need for belonging and relatedness in order to feel socially connected (Cho 2015).

With increasing remote work arrangements due to a shift in working models across industries and unforeseen crises such as the COVID-19 pandemic, individuals are subjected to the risk of social isolation (Rai 2020). In times of social isolation, using communication technology may help individuals to stay socially connected. However, this socio-technical circuit leads to compounded screen time and unprecedented consequences due to a constant permeation of technologically mediated communication in public, professional, and private activities. Consequently, perceived digital overuse can be observed as an emerging social issue and is defined as a widespread, but less pathological notion of feeling overwhelmed by communication content and connections (Gui and Büchi 2019). Affected individuals can experience negative consequences on both a private and professional level and experience their overall well-being to be impaired (Gui et al. 2017). Eventually, perceived digital overuse can lead to an IT addiction which is a psychological state of maladaptive dependency on IT use manifested through obsessive-compulsive patterns (Zhang et al. 2014).

To mitigate perceived digital overuse and its negative consequences on individuals' well-being, "digital detox" has been introduced as a possible counteragent (Gui and Büchi 2019). This popularised term can be defined as a periodic disconnection from social or online media as well as strategies which help reducing digital media involvement (Syvertsen and Enli 2019). For instance, this practice involves the planned limitation of screen time in order to reduce communication overload and digital overuse. Digital detox does not aim to object communication technology use altogether but rather supports awareness and provides strategies for a more intentional use of technology (Syvertsen and Enli 2019).

When conducting a digital detox, individuals may perceive a positive increase in overall well-being due to reduced perceived digital overuse (Gui and Büchi 2019; Syvertsen and Enli 2019). However, digital detox might also lead to less social connectedness caused by the absence of social media and communication technology usage leading to an increased feeling of loneliness (Cho 2015). Especially in case of social isolation, the communication technology, for some people, might be the main access point to their social relationships. Consequently, it remains to be investigated how the decrease in social connectedness during a digital detox affect individual well-being. Based on this argument, this research-in-progress paper focuses on the following research question:

*RQ: How do periods of digital detox impact perceived social connectedness?*

To answer this question, we outline a mixed-method approach consisting of a quantitative experiment and qualitative interviews. This will allow us to both measure the impact of digital detox periods on social connectedness, and subsequently, to delve the underlying causes we gain from the experiment. Our research is set out to provide important insights on the outcomes and consequences of digital detox, especially in times of social isolation. Digital detox, which is a relatively new phenomenon in IS research, will be scrutinised from an original angle, namely its possible negative effects on social connectedness. Moreover, practitioners such as professionals but also organisations may draw value from this study to better organise remote work practices and to build environments that prevent digital overuse but ensure social connectedness.

# 2   Background

## 2.1   Social Connectedness through Technology Usage

IT systems, including communication technology, are embedded in individuals' private and professional life (Techatassanasoontorn and Thaiprasert 2013). Those systems enable digitally embedded social connections. Individuals have the possibility to connect locally, with people in geographical proximity, across great distances, or completely independent of physical time and space (Chayko 2014). Purely digital socialising activities may as well generate feelings of warmth, belonging and excitement (Chmiel et al. 2011). Consequently, an individual's need for social connectedness can be satisfied and loneliness can be reduced (Chmiel et al. 2011). Social connectedness, in. this regard, can be defined as a short-term experience of relatedness and belonging. It can be evoked by quantitative and qualitative social appraisals and relationship salience. (van Bel et al. 2009). The latter emerges when thinking of others and may create the feeling of being together without physical contact (van Baren et al. 2004). Social connectedness can be found on two levels: the overall level and the individual level. On the overall level, individuals feel socially connected with their whole social network while on the individual level, individuals feel socially connected to one specific person (van Bel et al. 2009).

The counterpart to this is social isolation. Perceiving it mirrors a feeling of loneliness and is defined as a social pain evolving as a signal that an individual's connections to other people are weakening or are completely lost (Cacioppo and Hawkley 2009). Fear of social isolation spurs individuals' motivation to repair and maintain connections to others that are needed for their mental health and well-being. Research indicates that social isolation is a risk factor to i.e. poorer overall cognitive performance, increased negativity and depression (Cacioppo and Hawkley 2009). A study by Cho (2015) showed that intentional smartphone usage for connecting with other can help to reduce social isolation and increase social connectedness. Technology use, however, may also entail negative effects with regard to individual well-being. Inadvertent consequences, for instance, may be evoked when using technologies to an excessively high degree.

## 2.2 Perceived Digital Overuse

Research so far has mainly focused on Internet addiction when speaking of problematic, compulsive, or excessive Internet use which includes pathologic usage (Young 1998). However, perceiving general digital overuse is an emerging social issue which is less harmful but much more common (Gui and Büchi 2019). Since the Internet and social media are ubiquitous, constant availability has formed a new social standard. Features like push notifications trigger availability anytime and anywhere while often interrupts other ongoing activities (Büchi et al. 2019). However, when individuals do not respond in an appropriate time, they might be viewed as socially nonresponsive (Stephens et al. 2017). At the same time, individuals want to be always connected due to the so-called fear of missing out (FOMO) even though they feel overwhelmed. FOMO describes the fear of missing out important information or social interactions which are constantly happen online. As a consequent, FOMO leads to people using technology more often (Przybylski et al. 2013).

Perceiving the described feeling of being overwhelmed and cognitively overloaded can be described as perceived digital overuse (Büchi et al. 2019). Digital overuse is not a normative top-down perspective about what is "excessive" or "too much", it is rather an individual problematic feeling when using technology too often. Büchi et al. (2019) derive therefore the following definition: "Digital overuse is [...] a general and broad latent phenomenon that occurs when everyday Internet use surpasses an individual standard or vague sense of a personal optimum" (p.2). Since perceived digital overuse goes along with feeling stressed out and overwhelmed, affected individuals tried to find a way to minimise perceived digital overuse. A trend has raised where people start reducing their screen time of different devices (smartphone, tablet, computer) in order to mitigate stress and to focus on the offline world (Syvertsen and Enli 2019).

## 2.3 Digital Detox

Even though the concept of digital detox is relatively new, the usage of the term has rapidly increased (Syvertsen and Enli 2019). In general, detox defines a process in which one abstains from unhealthy objects (Basu 2019). In the year 2013, the term digital detox was introduced to the Oxford dictionary with the label: "A period of time during which a person refrains from using electronic devices such as smartphones or computers, regarded as an opportunity to reduce stress or focus on social interaction on the physical world"[1]. However, in scholarly literature, there is no accepted and defined notion at the time of this writing. Different terms are used to explain the same phenomenon like "digital diet" or "media diet" (Andersen et al. 2016). Still, the term "digital detox" has prevailed and is so far the most used term (e.g. Syvertsen & Enli, 2019; Wilcockson, Osborne, & Ellis, 2019). Syvertsen and Enli (2019) define digital detox as "[...] efforts to take a break from online or digital media for a longer or shorter period, as well as other efforts to restrict the use of smartphones and digital tools" (p. 2). The reasons for individuals practicing digital detox are to not being reachable for a while or to have more time for other things (Anrijs et al. 2018). Digital detox is influenced by a variety of inspirations and motives (Syvertsen and Enli 2019). One motive for practicing digital detox is based on the concept of self-optimisation. By that, individuals use technologies and software more efficiently in order to reduce digital distraction. Other motives are notions of balance and mindfulness. The overall intention of practicing digital detox is to avoid distractions that come from always being connected with the overall goal to reduce stress caused by digital overuse (Anrijs et al. 2018; Basu 2019).

---

[1] https://www.lexico.com/en/definition/digital_detox (last access: 09/08/2020)

To investigate if conducting a digital detox leads to the promised effect of reducing stress, Anrijs et al. (2018) conducted a study to investigate whether a smartphone digital detox effectively decreases stress in the short-term when individuals feel a perceived digital overuse. For that purpose, the authors monitored participants for two weeks. In the first week, the participants were allowed to use their smartphone like they normally do. In the second week, the participants practiced digital detox. To control the smartphone usage, the authors developed an application to track the screen time. Their results showed that physiological stress was reduced. Hence, prior research indicates that digital detox may be an effective coping mechanism for people who experience stress based on digital overuse.

Nonetheless, technology is not only a factor for stress, but also a tool for stress relief when feeling lonely or social isolated (Cho 2015). Its usage can help to evoke a feeling of social connectedness (van Bel et al. 2009). Therefore, the question arises whether digital detox in times of social isolation is still an appropriate concept or whether it rather causes more stress and harm caused by a reduced feeling of social connectedness.

## 3 Research Design

To answer the proposed research question comprehensively, a mixed method approach will be outlined including an experiment and qualitative interviews (Creswell et al. 2003). In doing so, both methods will mutually enrich one another by complementing quantitative and qualitative findings. The mixed method approach was chosen since the research in the field of digital detox is scarce. Thus, the results will not only be based on experimental data but also on real world experiences helping to gain a wide-range understanding of digital detox. While the experimental data will help to understand underlying processes and to build a theoretical understanding of digital detox, the interviews will link the research to real world impacts.

### 3.1 Experimental Setting

The study will commence with an experiment in order to answer the research question (*How does digital detox impact individuals' well-being in times of social isolation?*) on a quantitative level. The study for this phase will be carried out as a labour experiment. A pretest of our data collection instruments will be carried out to control for individual differences within two subgroups, and therefore, increase the precision of measurement. The requirements for the selection of participants are experience in working remotely or in a home office setting for at least six months. They should have at least a bachelor's degree and their occupation should fit the description of a knowledge worker, that is, for example, programmers, physicians, architects, engineers, or scientists. Their remote work experience and habits, which will be evaluated by a short self-assessment, must have confronted them with everyday usage of technologies in private and professional manner.

Since the goal is to examine the impact of conducting digital detox in times of high perceived social isolation, a feeling of being socially isolated needs to be provoked (Figure 1). Therefore, participants cannot be in touch with people from the offline world for two hours prior to the experiment. By that, a feeling of being lonely might occur. This setting will be the replicated for all participants. In order to measure the effect of digital detox, the participants will be randomly allocated in one of two groups. While being socially isolated, group A has to conduct a digital detox. For each group, we will canvass for at least 36 participants. The usage of any device is prohibited including the usage of their own smartphone. In the contrary, group B is allowed to use their smartphone during the whole time of the experiment. During the time of the experiment, all participants have the chance to entertain themselves with other activities like reading or drawing. Again, group A can additionally use their smartphone for entertainment purposes.

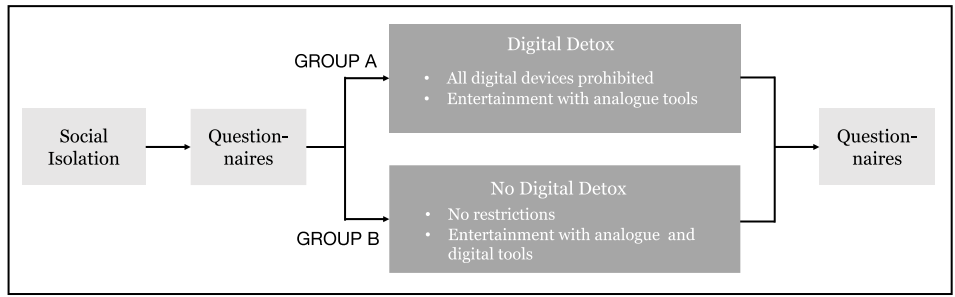

Figure 1. Experimental Setting

To test whether digital detox has reduced a feeling of social connectedness, all participants have to fill out different measures at the beginning and the end of the experiment. Table 1 depicts a preliminary questionnaire, including the purpose of the questionnaire, an example question and answer options.

| Questionnaire | Purpose | Example Question | Answer Option |
|---|---|---|---|
| Technostress (Maier et al. 2018) | Measuring whether feeling of technostress has been reduced | *I have to be always available due to this technology.* | 5-point Likert scale of agreement |
| Mobile Phone Craving (De-Sola et al. 2017) | Measuring whether digital detox leads to smartphone craving | *I would crave my smartphone if I wanted to turn it on right now and could not or would not be allowed to* | 10-point Likert scale of agreement |
| Perceived Digital Overuse (Gui and Büchi 2019) | Controlling how perceived digital overuse moderates the other factors | *When I use the Internet, I lose time for more important things* | 5-point Likert scale of agreement |
| Social Isolation (Zavaleta et al. 2017) | Controlling the success of the experimental setting | *I miss having people around* | 5-point Likert scale of agreement |
| Social Connectedness (van Bel et al. 2009) | Measuring whether feeling of social connectedness has been reduced | *Even when we are not in each other's company, I often feel "together" with people in my social network somehow* | 7-point Likert Scale of agreement |
| Fear of Missing out | Testing if digital detox influences fear of missing out | I fear others have more rewarding experience than me | 5-point Likert scale of agreement |
| Overall well-being | Testing whether digital detox influenced the overall well-being positively or negatively | *I feel useful and needed* | 5-point Likert scale of agreement |

*Table 1. Quantitative measures for the experimental design.*

## 3.2　Qualitative Interviews

As a second step of the study, qualitative interviews will be conducted with the aim to get deeper insights but also to see how the results from the experiment can be applied in a real-world setting (Braun and Clarke 2006). The goal is to answer the research question additionally on a qualitative level. An interview guideline is developed by deducting questions from the theoretical constructs from table 1. Three concepts based on the theoretical background can be identified to address the research question (Table 2). By that, the foundation for the interview guideline is created.

The study is set out to conduct a minimum number of twelve semi-structured interviews. Those allow to address all relevant aspects in order to answer the research question, but also give room for new ideas and more insights into the topic (Myers and Newman 2007). Table 2 provides the basis for the interview guidelines.

| Concepts | Category |
|---|---|
| Concept A: Perceived Digital Overuse | Experience with technostress |
| | Role of smartphone & social media |
| | Experience with communication overload |
| | Experience with information overload |
| Concept B: Social Isolation | Knowledge about the pandemic COVID-19 |
| | Experience of social isolation |
| | Role of digital media during COVID-19 |
| Concept C: Digital Detox | Attitude toward digital detox |
| | Motives for digital detox |
| | Experience with digital detox |
| | Digital detox during COVID-19 |

Table 2. Relevant concepts for creating the interview guidelines.

## 4　Expected Results

Even though digital detox is an emerging catchword in popular culture, scholarship has yet to empirically strengthen the concept. Anrijs et al. (2018) showed that digital detox can reduce perceived stress when being practiced. However, using technologies can also have many benefits like feeling socially connected or reducing the feeling of being socially isolated (van Bel et al. 2009; Cho 2015). Therefore, we expect that practicing a digital detox in times of social isolation might also entail feeling of social disconnect during times of social isolation. Being isolated not only in the offline but also in the online world might evoke a feeling of loneliness, and the level of stress might increase which both can negatively impact overall well-being. The experimental setting might be improved in a way that social isolation is not only provoked by being alone for three hours without any offline social contact. However, the decision for this kind of setting was still made since it is inappropriate and unethical to socially isolate participants in their real life for days since their well-being can be seriously harmed by it (Cacioppo and Hawkley 2009). In order to control the feeling of isolation caused by the experimental setting, a questionnaire to measure such feeling is included for the start and the end of the experiment (Zavaleta et al. 2017). Conducting additional interviews will mitigate this limitation as participants may fall back on their own experiences of "social distancing" during the COVID-19 pandemic.

## 5　Conclusion and Next Steps

A research design was derived and developed in order to answer the research question on both a qualitative and a quantitative level accordingly. The experiment consists of established questionnaires which are known to measure the different items appropriately. The next steps for this study will be do conduct both the experiment and the interviews. Prospective results may help to understand whether digital detox practiced while being socially isolated reduces perceived digital overuse, and thus, impacts individuals' well-being positively.

Experimental data can help to enrich underlying processes and understanding of digital detox on a quantitative level. At the same time, real-world experiences can give in-depth insights into the experience of a digital detox when being socially isolated. This paper can contribute to theory and practice by developing guidelines for individual professionals and organisations practicing remote work arrangements. Implications of this research can help to understand in which case digital detox is a helpful tool to reduce stress and when it is not. At the same time, enterprises might profit from the deeper understanding of digital detox especially when employees work from home virtually. Enterprises can support employees by either suggesting digital detox or not in order to reduce stress and improve individuals' well-being.